\documentclass[twocolumn,english,showpacs]{revtex4}
\usepackage{times}
\usepackage[OT1]{fontenc}
\usepackage[latin1]{inputenc}
\usepackage{graphicx}
\usepackage{amssymb}

\makeatletter

\usepackage{amsmath}
\usepackage{graphicx}
\usepackage{amssymb}

\usepackage{graphicx}

\usepackage{babel}

\usepackage{babel}
\makeatother
\begin{document}

\title{Adiabatic cooling of Fermions in an optical lattice }

\author{P. B. Blakie and A. Bezett}

\affiliation{Department of Physics, University of Otago, P.O. Box 56, Dunedin,
New Zealand }

\date{\today{}}

\begin{abstract}
The entropy-temperature curves are calculated for non-interacting
fermions in a 3D optical lattice. These curves facilitate understanding
of how adiabatic changes in the lattice depth affect the temperature,
and we demonstrate regimes where the atomic sample can be significantly
heated or cooled. When the Fermi energy of the system is near the
location of a band gap the cooling regimes disappear for all temperatures
and the system can only be heated by lattice loading. For samples
with greater than one fermion per site we find that lattice loading
can lead to a large increase in the degeneracy of the system. We study
the scaling properties of the system in the degenerate regimes and
assess the effects of non-adiabatic loading. 
\end{abstract}


\maketitle

\section{Introduction}

Tremendous progress has been made in the preparation, control and
manipulation of Fermi gases in the degenerate regime \cite{DeMarco1999a,Schreck2001a,O'Hara2002a,Modugno2002a,Gupta2003a,Regal2003a,Cubizolle2003a}.
Such systems have many potential applications in the controlled study
of fermionic superfluidity and the production of ultra-cold molecules.
Another area of developing theoretical interest is in the physics
of fermions in optical lattices \cite{Hofstetter2002,Rabl2003,Viverit2004,Santos2004},
and initial experiments have already begun to examine the properties
of Fermi-gases (prepared as boson-fermion mixtures) in one-dimensional
optical lattices \cite{Modugno2003a,Ott2004a}. For Bose gases, optical
lattices have been used to demonstrate an impressive array of experiments
such as: quantum matter-wave engineering \cite{Orzel2001a,Greiner2002b};
the Mott-insulator quantum-phase transition \cite{Greiner2002a};
quantum entanglement \cite{Mandel2003a}; and coherent molecule production
\cite{Rom2004a}. It seems likely that a similar range of rich physics
lies ahead for fermions in optical lattices.

Many of the physical phenomenon that are suitable to experimental
investigation in optical lattices are sensitive to temperature and
it is therefore of great interest to understand how the temperature
of a quantum degenerate gas changes with lattice depth. Experimental
results by Kastberg \emph{et al.} \cite{Kastberg1995a} in 1995 showed
that loading laser cooled atoms into a three-dimensional optical lattice
caused the atoms to increase their temperature %
\footnote{In fact this study used adiabatic de-loading to reduce the temperature
of the constituent atoms.%
}. Recently one of us conducted a detailed thermodynamic study of bosonic
atoms in optical lattices \cite{Blakie2004a}. In that work we showed
that for sufficiently low initial temperatures a new regime would
be entered in which adiabatically ramping up the lattice depth would
have the desirable effect of cooling the system. The typical temperatures
at which Bose-Einstein condensates are produced lie well within this
cooling regime, and thus benefit from reduced thermal fluctuations
when adiabatically loaded into an optical lattice. In this paper we
examine how degenerate fermions are affected by adiabatic loading
into an optical lattice. In Fermi gases the lowest temperatures obtained
in experiments tend to be much higher than in Bose gas experiments.
It is therefore important to understand to what extent the introduction
of an optical lattice might affect the temperature, in particular
to determine in what regimes additional cooling can occur during lattice
loading.

The quintessential difference in the properties of degenerate fermions
from bosons is embodied by the Fermi energy - the energy that marks
the top of the Fermi-sea of occupied states (at $T=0$). The Fermi
energy sets a new energy scale that has no analog in boson systems
and plays a crucial role in determining the effect that lattice loading
has upon the system. We find that as the Fermi energy approaches a
band gap, the cooling regime vanishes and the system can only heat
with increasing lattice depth. However, we also find that when the
Fermi energy lies in the second band (when the average number of fermions
per site is greater than one) a cooling regime is re-established.
This cooling regime for the second band is accompanied by a large
amplification of degeneracy, i.e. adiabatically loading into the lattice
causes both $T$ and the ratio $T/T_{F}$ to decrease.

The results we present in this paper are obtained from a numerical
study of the thermodynamic properties of an ideal gas of fermions
in a 3D cubic lattice. We work with the grand canonical ensemble and
use the exact single particle eigenstates of the lattice to determine
the entropy-temperature curves for the system for various lattice
depths and filling factors. We develop analytic expressions for the
plateaus that develop in the entropy-temperature curves and characterize
a scaling relationship that holds for low temperatures and in deep
lattices. A \emph{fast loading} procedure is considered to ascertain
how robust our results are to non-adiabatic effects. The physics we
explore here will be relevant to current experiments, and many of
the predictions we make should be easily seen.

\section{Formalism}

\subsection{Single Particle Eigenstates}

We consider a cubic 3D optical lattice made from 3 independent (i.e.
non-interfering) sets of counter-propagating laser fields of wavelength
$\lambda$, giving rise to a potential of the form\begin{equation}
V_{{\textrm{Latt}}}(\mathbf{r})=\frac{V}{2}[\cos(2kx)+\cos(2ky)+\cos(2kz)],\label{eq:LattPot}\end{equation}
 where $k=2\pi/\lambda$ is the single photon wavevector, and $V$
is the lattice depth. We take the lattice to be of finite extent with
a total of $N_{s}$ sites, consisting of an equal number of sites
along each of the spatial directions with periodic boundary conditions.
The single particle energies \textbf{$\epsilon_{\mathbf{q}}$} are
determined by solving the Schr\"{o}dinger equation

\begin{equation}
\epsilon_{\mathbf{q}}\psi_{\mathbf{q}}(\mathbf{r})=\frac{\mathbf{p^{2}}}{2m}\psi_{\mathbf{q}}(\mathbf{r})+V_{{\textrm{Latt}}}(\mathbf{r})\psi_{\mathbf{q}}(\mathbf{r}),\label{eq:BlochState}\end{equation}
 for the Bloch states, $\psi_{\mathbf{q}}(\mathbf{r}),$ of the lattice.
For notational simplicity we choose to work in the extended zone scheme
where $\mathbf{q}$ specifies both the quasimomentum and band index
of the state under consideration %
\footnote{For a discussion of how the quantum numbers of quasimomentum and band
index are introduced we refer the reader to Ref. \cite{Mermin1976}%
}. By using the single photon recoil energy, $E_{R}=\hbar^{2}k^{2}/2m$,
as our unit of energy, the energy states of the system are completely
specified by the lattice depth $V$ and the number of lattice sites
$N_{s}$ (i.e. in recoil units $\epsilon_{\mathbf{q}}$ is independent
of $k$).

\begin{figure}
\includegraphics[%
  width=3.1in,
  keepaspectratio]{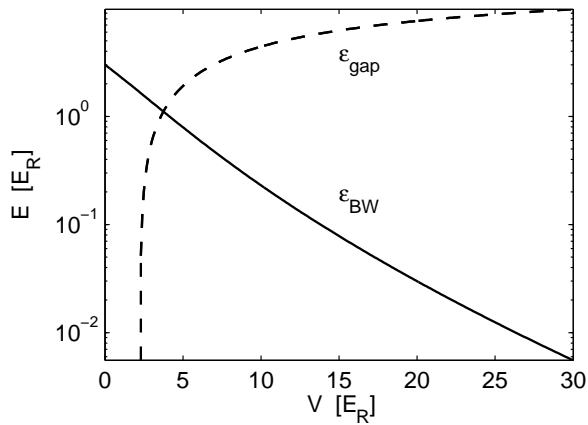}

\caption{\label{cap:Egapbw} The dependence of the energy gap ($\epsilon_{{\textrm{gap}}}$
dashed line) and ground band width ($\epsilon_{{\textrm{BW}}}$ solid
line) on the lattice depth (see the text).}
\end{figure}

For completeness we briefly review some important features of the
band structure of Eq. (\ref{eq:BlochState}) relevant to the thermodynamic
properties of the system. For sufficiently deep lattices an energy
gap, $\epsilon_{{\textrm{gap}}}$, will separate the ground and first
excited bands (see Fig. \ref{cap:Egapbw}). For the cubic lattice
we consider here, a finite gap appears at a lattice depth of $V\approx2E_{R}$
\footnote{The delay in appearance of the excitation spectrum gap until $V\approx2E_{R}$
is a property of the 3D band structure. In a 1D lattice a gap is present
for all depths $V>0$.%
} (marked by the vertical asymptote of the dashed line in Fig. \ref{cap:Egapbw}).
For lattice depths greater than this, the gap increases with lattice
depth. In forming the gap higher energy bands are shifted up-wards
in energy, and the ground band becomes compressed --- a feature characteristic
of the reduced tunneling between lattice sites. We refer to the energy
range over which the ground band extends as the (ground) band width,
$\epsilon_{{\textrm{BW}}}$. As is apparent in Fig. \ref{cap:Egapbw},
the ground band width decreases exponentially with $V$ , causing
the ground band to have an extremely high density of states for deep
lattices.

\subsection{Equilibrium Properties}

Our primary interest lies in understanding the process of adiabatically
loading a system of $N_{p}$ fermions into a lattice. Under the assumption
of adiabaticity the entropy remains constant throughout this process
and the most useful information can be obtained from knowing how the
entropy depends on the other parameters of the system. In the thermodynamic
limit, where $N_{s}\rightarrow\infty$ and $N_{p}\rightarrow\infty$
while the filling factor $n\equiv N_{p}/N_{s}$ remains constant,
the entropy per particle is completely specified by the intensive
parameters $T,\, V,\, n$. The calculations we present in this paper
are for finite size systems, that are sufficiently large to approximate
the thermodynamic limit. We would like to emphasize at this point
the remarkable fact that $V$ is an adjustable parameter in optical
lattice experiments, in contrast to solid state systems where the
lattice parameters are immutable.

The entropy is determined as follows: The single particle spectrum
$\{\epsilon_{\mathbf{q}}\}$ of the lattice is calculated for given
values of $N_{s}$ and $V$. We then determine the thermodynamic properties
of the lattice with $N_{p}$ fermions in the grand canonical ensemble,
for which we calculate the partition function $\mathcal{Z}$\begin{equation}
\log\mathcal{Z}=\sum_{\mathbf{q}}\log\left(1+e^{-\beta(\epsilon_{\mathbf{q}}-\mu)}\right),\label{eq:GrandPot}\end{equation}
 where $\mu$ is found by ensuring particle conservation. The entropy
of the system can then be expressed as\begin{equation}
S=k_{B}\left(\log\mathcal{Z}+\beta E-\mu\beta N_{p}\right),\label{eq:Entropy}\end{equation}
 where $\beta=1/k_{B}T$, and $E=-\partial\ln\mathcal{Z}/\partial\beta$
is the mean energy.

\subsubsection{Multiple components}

In most current experiments mixtures of Fermi gases in different internal
states are studied. This is required because s-wave elastic collisions,
needed for re-equilibration, are prohibited by the Pauli principle
for spin polarized samples %
\footnote{This also means that a single component Fermi gas is quite well described
by a non-interacting theory.%
}. The theory we present here is for the spin polarized case, but is
trivially extensible to multiple components if the lattice potential
is spin independent and the number of atoms in each component is the
same: in this case all extensive parameters are doubled (e.g. $\{ E,S\}$)
and intensive parameters (e.g. $\{ T,\mu\}$) remain the same. The
inclusion of interaction effects, which will be important in the multiple
component case, is beyond the scope of this paper.

\section{Results}

\subsection{Effect of lattice loading on Fermi-gas temperature}

\begin{figure*}
\includegraphics[%
  width=7in,
  keepaspectratio]{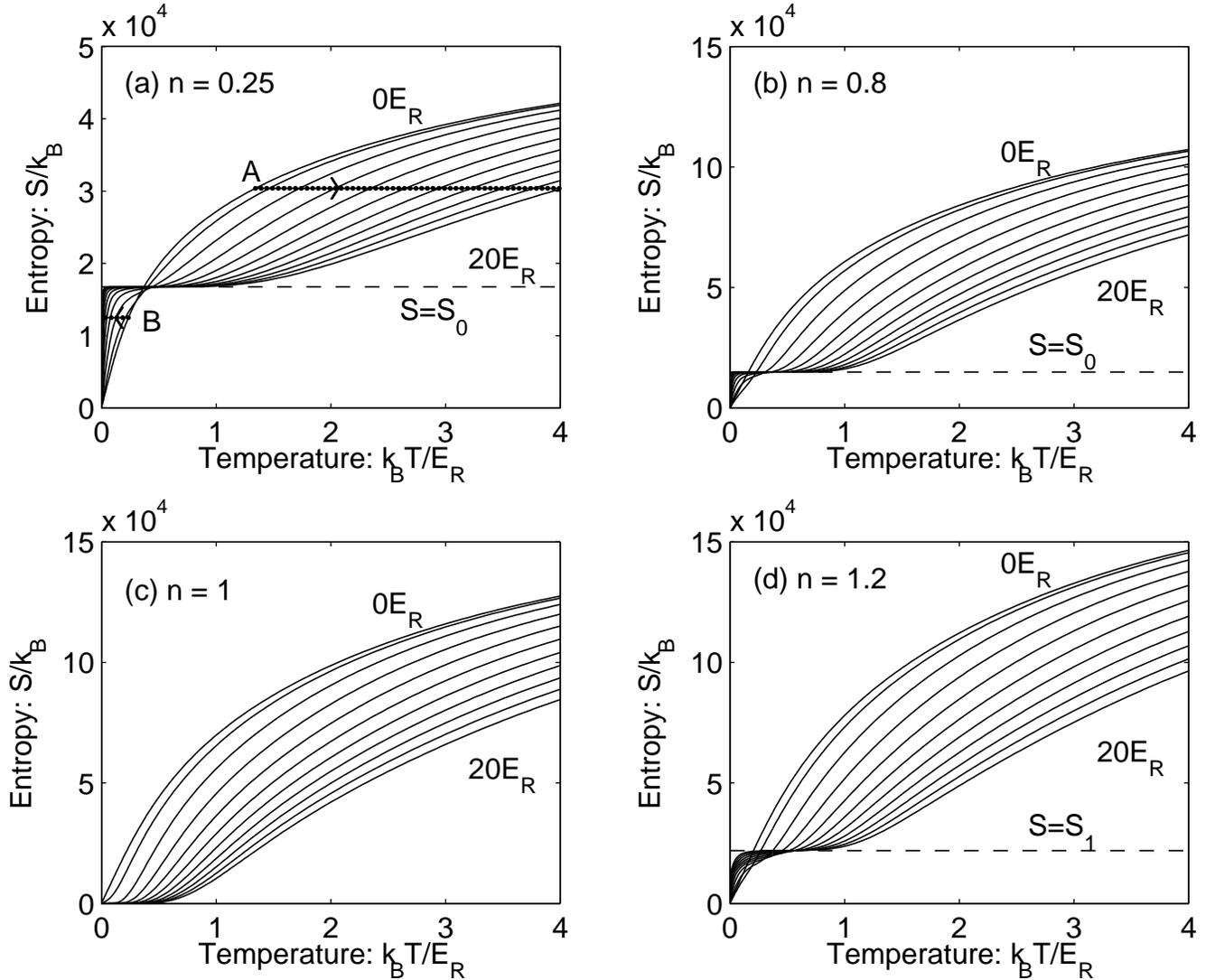}

\caption{\label{FIG:ST1} Entropy versus temperature curves for a $N_{s}\approx3\times10^{4}$
site cubic lattice, at various depths $V=0$ to $20E_{R}$ (with a
spacing of $2E_{R}$ between each curve). Filling factors used are
(a) $n=0.25,$ (b) $n=0.8,$ (c) $n=1.0,$ and (d) $n=1.2$. The entropy
plateau is shown as a dashed line. Dotted line marked $A$ shows a
path along which adiabatic loading into the lattice causes the temperature
to increase. Dotted line marked $B$ shows a path along which adiabatic
loading into the lattice causes the temperature to decrease. }
\end{figure*}

In Fig. \ref{FIG:ST1} we show entropy-temperature curves for various
lattice depths and filling factors $n$. These curves have been calculated
for a lattice with $31$ lattice sites along each spatial dimension,
i.e. $N_{s}\approx3\times10^{4}$.

A general feature of these curves is the distinct separation of regions
where adiabatic loading causes the temperature of the sample to increase
or decrease, which we will refer to as the regions of heating and
cooling respectively. These regions are separated by a value of entropy
at which the curves plateau - a feature that is more prominent in
the curves for larger lattice depths. This plateau entropy is indicated
by a horizontal dashed line and is discussed below. For the case of
unit filling factor shown in Fig. \ref{FIG:ST1}(c), this plateau
occurs at $S=0$, and only a heating region is observed.

We now explicitly demonstrate the temperature changes that occur during
adiabatic loading using two possible adiabatic processes labeled $A$
and $B$, and marked as dashed-dot lines in Fig. \ref{FIG:ST1}(a).
Process $A$ begins with a gas of free particles in a state with an
entropy value lying above the plateau entropy. As the gas is loaded
into the lattice the process line indicates that the temperature increases
rapidly with the lattice depth. Conversely process $B$ begins with
a gas of free particles in a state with entropy below the plateau.
For this case adiabatic lattice loading causes a rapid decrease in
temperature. This behavior can be qualitatively understood in terms
of the modifications the lattice makes to the energy states of the
system. As is apparent in Fig. \ref{cap:Egapbw}, the ground band
rapidly flattens for increasing lattice depth causing the density
of states to be more densely compressed at lower energies. Thus in
the lattice all these states can be occupied at a much lower temperature
than for the free particle case. As we discuss below, $S_{0}$ is
the maximum entropy available from only accessing states of the lowest
band. If $S<S_{0}$, the temperature of the system must decrease with
increasing lattice depth to remain at constant entropy. Alternatively,
for $S>S_{0}$ the occupation of states in higher bands is important,
and as the lattice depth and hence $\epsilon_{{\textrm{gap}}}$ increases,
the temperature must increase for these excited states to remain accessible.

\subsection{Fermi-gas degeneracy}

\begin{figure*}
\includegraphics[%
  width=7in,
  keepaspectratio]{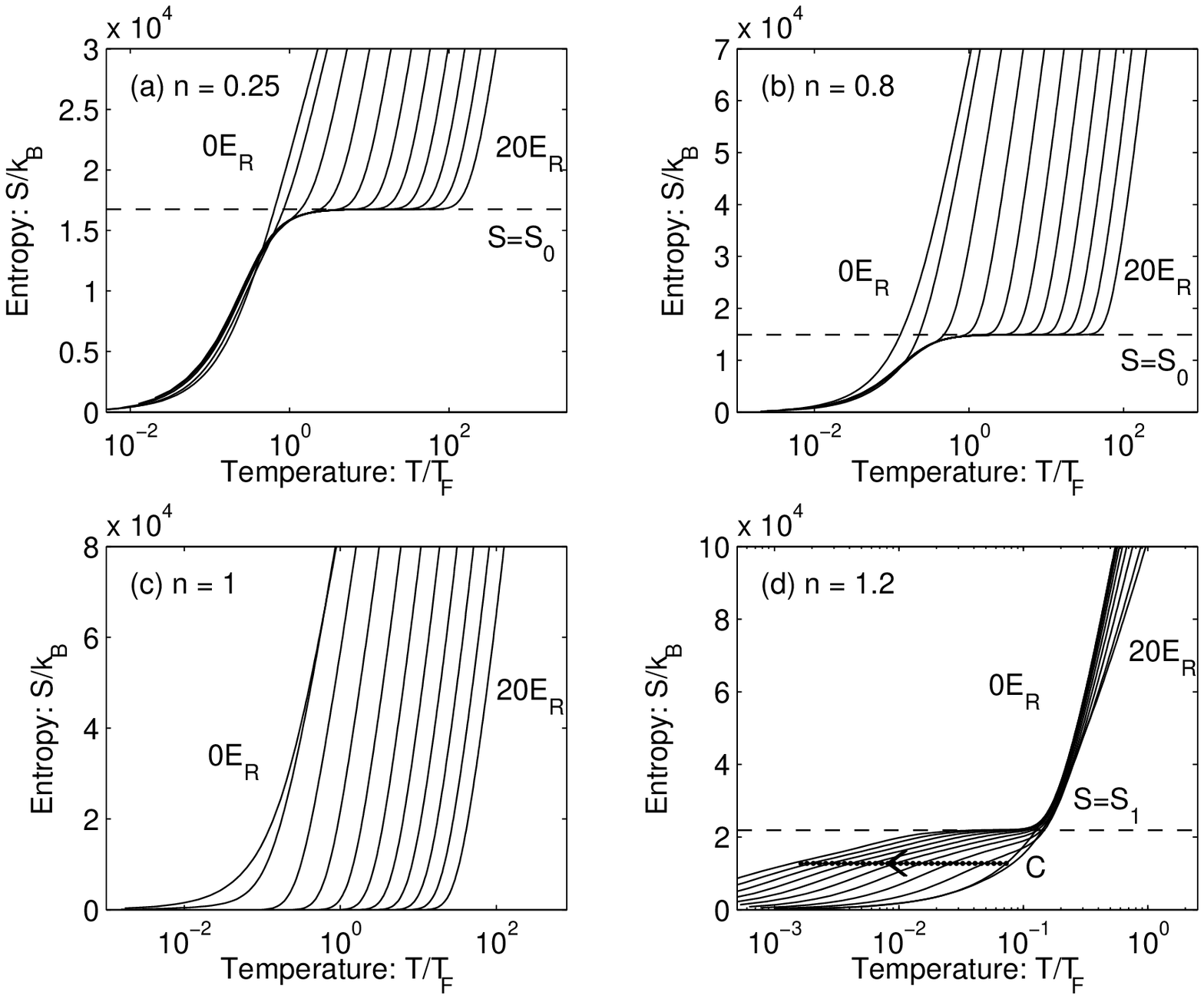}

\caption{\label{FIG:ST2} Entropy versus temperature over Fermi temperature
curves for the same cases considered in Fig. \ref{FIG:ST1} (see that
figure for details of parameters used). The entropy plateau is shown
as a dashed line. The dotted line marked $C$ shows a path along which
adiabatic loading into the lattice causes the ratio of the temperature
to the Fermi temperature to decrease.}
\end{figure*}

In addition to the effect that lattice loading has on the temperature
of a Fermi-gas, it is of considerable interest to understand how the
ratio of temperature to the Fermi temperature ($T_{F}$)%
\footnote{The Fermi temperature is given by $T_{F}=\epsilon_{F}/k_{B},$ where
$\epsilon_{F}$ (the Fermi energy) is the energy of the highest occupied
single particle state for the system at $T=0K$.%
} changes. Indeed, the ratio $T/T_{F}$ is the standard figure of merit
used to quantify the degeneracy of dilute Fermi gases. In Fig. \ref{FIG:ST2}
we show how $T/T_{F}$ changes with adiabatic lattice loading for
the same parameters used in Fig. \ref{FIG:ST1}. In Fig. \ref{FIG:ST2}
(a)-(b) the same general behavior is seen: Below the entropy plateau
where cooling is observed (see Fig. \ref{FIG:ST1} (a)-(b)), the ratio
of $T/T_{F}$ remains approximately constant, so that there is little
change in the degeneracy of the gas. Above the entropy plateau where
heating was observed, the ratio of $T/T_{F}$ rapidly increases, so
that in this regime the gas will rapidly become non-degenerate as
it is loaded into the lattice. For the unit filling case (Fig. \ref{FIG:ST2}
(c)), there is no cooling regime, and heating is accompanied by a
rapid increase in $T/T_{F}$ for all initial conditions of the gas.
In Fig. \ref{FIG:ST2} (d), where the filling factor is $n=1.2$,
rather different behavior is seen: In the cooling regime, the ratio
of $T/T_{F}$ is rapidly suppressed as the temperature decreases,
e.g. see the dotted line marked $C$ in Fig. \ref{FIG:ST2} (d). This
most desirable behavior could be used for example to prepare a Fermi-gas
into a highly degenerate state where the BCS transition might be observable.
We also note that for the same parameters but in the heating regime
the ratio $T/T_{F}$ remains relatively constant. 

We can give a simple explanation for the behavior of $T/T_{F}$. For
the three cases considered in Fig. \ref{FIG:ST2} (a)-(c) the Fermi
energy lies within or at the top of the first band of energy states.
As shown in Fig. \ref{cap:Egapbw} the width of the ground band ($\epsilon_{{\rm BW}}$)
decreases rapidly with lattice depth. Because the number of states
contained in each band is constant (given by the number of lattice
sites) both the Fermi energy and $T_{F}$ scale identically to $\epsilon_{{\rm BW}}$,
thus will rapidly decrease with lattice depth. In the cooling regime,
the temperature scales in the same manner as $\epsilon_{{\rm BW}}$
(see Sec. \ref{sub:Tight-binding-limit-with} below and Fig. \ref{cap:ScalingTfig}),
and thus the ratio $T/T_{F}$ remains approximately constant. In the
heating regime $T$ increases slowly, while the ratio $T/T_{F}$ increases
rapidly with lattice depth (on account of $T_{F}$ becoming small).

For the case considered in Fig. \ref{FIG:ST2} (d) the filling factor
satisfies $n>1$ and the Fermi energy lies in the second band. As
the lattice depth increases the Fermi energy and $T_{F}$ now scale
like $\epsilon_{{\rm gap}}$, i.e. slowly increases with lattice depth
(see Fig. \ref{cap:Egapbw}). Thus in the regime where the temperature
decreases, the ratio $T/T_{F}$ must become smaller. We note that
the temperature reduction occurs because the width of the second band
decreases with lattice depth.

\subsection{Entropy plateau\label{sub:Entropy-Plateau}}

In Figs. \ref{FIG:ST1}(a) and (b) a horizontal plateau (at the level
marked by the dashed lines) is common to the entropy-temperature curves
for larger lattice depths ($V\gtrsim8E_{R}$). This occurs because
for these lattices, the energy range over which the ground band extends
is small compared to the energy gap to the excited band, and there
is a large temperature range over which states in the excited bands
are unaccessible, yet all the ground band states are uniformly occupied.
The entropy value indicated by the dashed line in Figs. \ref{FIG:ST1}(a)
and (b) corresponds to the total number of $N_{p}$-particle states
in the ground band. Since the number of single particle energy states
in the ground band is equal to the number of lattice sites, the total
number of available $N_{p}$-particle states ($\Omega_{0}$) is given
by $\Omega_{0}=N_{s}!/[N_{p}!(N_{s}-N_{p})!]$ (valid for $N_{p}\leq N_{s}$).
The associated entropy $S_{0}=k_{B}\log\Omega_{0}$, which we shall
refer to as the plateau entropy, can be evaluated using Sterling's
approximation \begin{eqnarray}
S_{0} & \simeq & k_{B}\Big[N_{p}\log N_{p}+N_{s}\log N_{s}\label{eq:Splateau}\\
 &  & -(N_{s}-N_{p})\log\left(N_{s}-N_{p}\right)\Big],\nonumber \end{eqnarray}
 the validity condition for this result is that $1\ll N_{s}\ll N_{p}$.
An important case for which the above approximation is invalid is
for $N_{p}=N_{s},$ i.e. we have a filling factor of $n=1$, where
$S_{0}=0$. This case corresponds to the unit filling factor result
shown in Fig. \ref{FIG:ST1}(c) where, as a result of the entropy
plateau occurring at $S=0,$ only a heating region is observed.

Similar entropy plateaus are observed for greater than unit filling
($N_{p}>N_{s}$), e.g. as is seen in Fig. \ref{FIG:ST1}(d). For fermions
such high filling factors necessarily means that higher bands are
occupied, and in general the precise details of these higher plateaus
will depend on the particle band structure of the lattice. E.g in
the lattice we consider here (\ref{eq:LattPot}) there are three degenerate
first excited bands, that contain a total of $3N_{s}$ single particle
states. Because the first band is fully occupied only $N_{p}^{'}=N_{p}-N_{s}$
particles are available to occupy the excited band, and so the total
number of available states is found according to the ground band result
(\ref{eq:Splateau}), but with the substitutions $N_{s}\rightarrow3N_{s}$
and $N_{p}\rightarrow N_{p}-N_{s}$. This result is shown as the dashed
horizontal in Fig. \ref{FIG:ST1}(d) labelled as $S_{1}$.

The suppression of the plateaus at specific integer filling factors
(e.g. $n=1$ for $S_{0}\rightarrow0$ and $n=4$ for $S_{1}\rightarrow0$)
corresponds to the Fermi-energy of the system approaching a band gap.
Whenever this occurs it means that all the states below the gap are
occupied at $T=0$ and excitations in the system require the promotion
of particles into the excited band (above the gap). As all band gaps
increase in size with lattice depth, the temperature of the system
must increase for the entropy to remain constant. Thus in regimes
where the Fermi-energy lies at a band gap the system only exhibits
heating with increasing lattice depth (e.g. see Fig. \ref{FIG:ST1}(c)).

\subsection{Scaling: Tight-binding limit at low temperatures and filling factors
\label{sub:Tight-binding-limit-with}}

Here we give limiting results for the entropy-temperature curves.

As discussed in Sec. \ref{sub:Entropy-Plateau}, when $N_{p}<N_{s}$
and the temperature is sufficiently low that $S<S_{0}$, then only
single particle states within the ground band are accessible to the
system. In addition when the tight-binding description is applicable
for the initial and final states of an adiabatic process, the initial
and final thermodynamic variables are related by a scaling transformation.

In the tight-binding regime, which is good approximation for $V\gtrsim4E_{R}$,
the ground band dispersion relation takes the form \begin{equation}
\epsilon_{{\textrm{TB}}}(\mathbf{q})=-\frac{\epsilon_{{\textrm{BW}}}}{6}\sum_{j=\{ x,y,z\}}\cos(q_{j}a),\label{eq:TB}\end{equation}
 where $a=\pi/\lambda$ is the lattice period, the ground bandwidth
$\epsilon_{{\textrm{BW}}}$ has already been introduced (e.g. see
Fig. \ref{cap:Egapbw}), and the wavevector $\mathbf{q}$ is restricted
to the first Brillouin zone. We refer the reader to Refs. \cite{Jaksch1998a,Oosten2001a}
for more details on the tight-binding approximation.

To illustrate the scaling transformation we consider an initial system
in equilibrium with entropy $S<S_{0}$, in lattice of depth $V_{i}$
sufficiently large enough for tight-binding expression (\ref{eq:TB})
to provide an accurate description of the ground band energy states.
If an adiabatic process is used to take the system to some final state
at lattice depth $V_{f}$ (also in the tight-binding regime) it is
easily shown that the macroscopic parameters of the initial and final
states are related as \begin{equation}
X_{f}=\alpha X_{i},\label{eq:scalingrelation}\end{equation}
 where $X=\{ E$, $T$, or $\mu\}$, and the scaling parameter $\alpha=(\epsilon_{{\textrm{BW}}})_{f}/(\epsilon_{{\textrm{BW}}})_{i}$
is given by the ratio of the final and initial bandwidths. The requirement
that the initial and final states are in the tight-binding regime
is because the single particle states are then related as $\left(\epsilon_{{\textrm{TB}}}(\mathbf{q})\right)_{f}=\alpha\left(\epsilon_{{\textrm{TB}}}(\mathbf{q})\right)_{i}$,
which is essential for (\ref{eq:scalingrelation}) to hold.

This type of scaling suggests that the occupations of the single particle
levels are unchanged during the change in lattice: the products $\beta\epsilon_{{\textrm{TB}}}(\mathbf{q})$
and $\beta\mu$ are independent of $V$, so the Fermi distribution,
$f_{F}(\mathbf{q})=[\exp(\beta\epsilon_{{\textrm{TB}}}(\mathbf{q})-\beta\mu)+1]^{-1}$
will also be independent of $V$. This suggests that being adiabatic
in this regime will not require redistribution through collisions
and may allow the lattice depth to be changed more rapidly. 

\begin{figure}
\includegraphics[%
  width=3.5in,
  keepaspectratio]{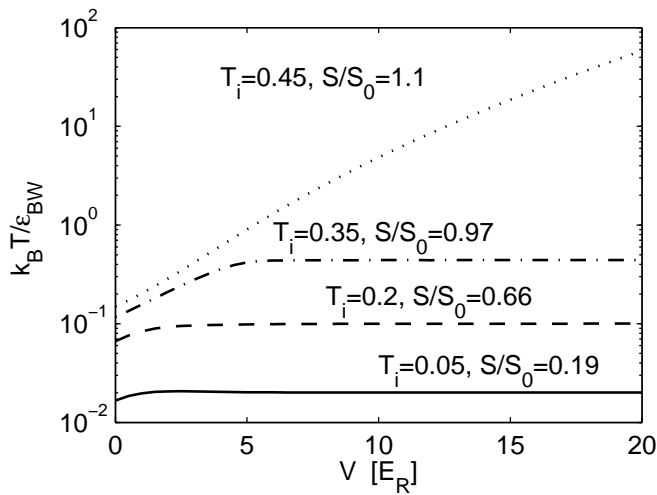}

\caption{\label{cap:ScalingTfig} Scaled temperature change during adiabatic
lattice loading. Ratio of temperature to bandwidth for various lattice
depths along adiabatic contours for several values of entropy. Initial
temperature (at $V=0$) and ratio of entropy to the plateau entropy
is indicated for each curve. Results are indicated for the case of
filling factor $n=0.25$.}
\end{figure}

To confirm the scaling predictions, in Fig. \ref{cap:ScalingTfig}
we plot the ratio of the temperature to ground bandwidth as a function
of lattice depth along contours of constant entropy, (e.g. how $k_{B}T/\epsilon_{{\textrm{BW}}}$
varies along the process curves labelled $A$ and $B$ in Fig. \ref{FIG:ST1}(a)).
In regimes where the scaling relationship (\ref{eq:scalingrelation})
holds true, the ratio $k_{B}T/\epsilon_{{\textrm{BW}}}$ should be
constant (independent of $V$). In Fig. \ref{cap:ScalingTfig} this
is clearly observed for initial entropies less than $S_{0}$ and lattice
depths $V\gtrsim5E_{R}$. For $S>S_{0}$ single particle states of
higher bands necessarily play an important role in the thermodynamic
state of the system, and the scaling transformation clearly does not
hold at any lattice depth, as is seen in the dotted curve in Fig.
\ref{cap:ScalingTfig}. For this case as the lattice depth increases
the cooling effect of the ground band compression is offset by the
f particles in the excited band that are lifted to larger energies
as the gap ($\epsilon_{{\textrm{gap}}}$) grows (see Fig. \ref{cap:Egapbw}).

\subsection{Adiabaticity\label{sub:Adiabaticity}}

\begin{figure}
\includegraphics[%
  width=3.6in,
  keepaspectratio]{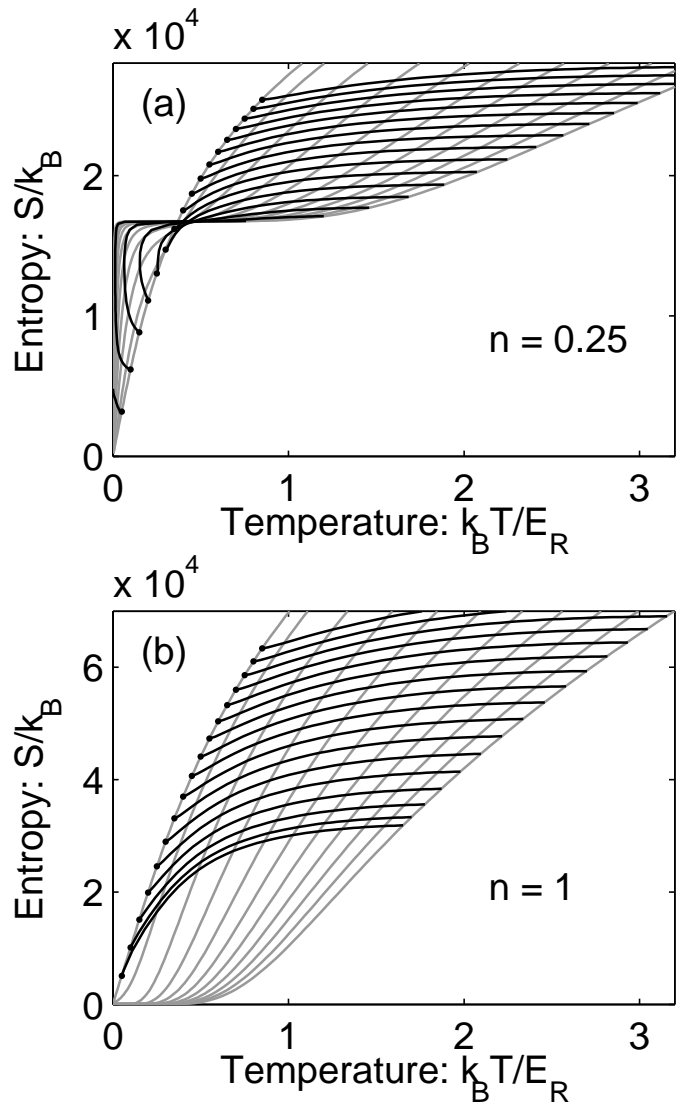}

\caption{\label{FIG:nonadiab} Fast lattice loading of a $N_{s}\approx3\times10^{4}$
site cubic lattice, with filling factors of (a) $n=0.25$ and (b)
$n=1$. Dark solid lines indicate fast loading curves (see text).
The initial $V=0$ state for each of this curves is indicated with
a dot. The lattice depth on these curves can be determined from their
intercept with the equilibrium entropy versus temperature curves (gray
solid lines), which are described in Figs. \ref{FIG:ST1}(a) and (c)
for $n=0.25$ and $n=1$ respectively.}
\end{figure}

Finally we note that interactions between particles are essential
for establishing equilibrium in the system, and understanding this
in detail will be necessary to determine the timescale for adiabatic
loading. In general this requirement is difficult to assess, and in
systems where there is an additional external potential it seems that
the adiabaticity requirements will likely be dominated by the process
of atom transport within the lattice to keep the chemical potential
uniform, though recent proposals have suggested ways of reducing this
problem \cite{Sklarzr2002a} for Bose systems. A study of the effects
of interactions or inhomogeneous potentials is beyond the scope of
this work, however it is useful to assess the degree to which non-adiabatic
loading would cause heating in the system. We consider lattice loading
on a time scale fast compared to the typical collision time between
atoms, yet slow enough to be quantum mechanically adiabatic with respect
to the single particle states. This latter requirement excludes changing
the lattice so fast that band excitations are induced, and it has
been shown that in practice this condition can be satisfied on very
short time scales \cite{Denschlag2002a}. We will refer to this type
of loading as fast lattice loading, to distinguish it from the fully
adiabatic loading we have been considering thus far.

To simulate the fast lattice loading we take the system to be initially
in equilibrium at temperature $T_{i}$ for zero lattice depth. For
the final lattice depth we fast load into, we map the initial single
particle distribution onto their equivalent states in the final lattice,
and calculate the total energy for this final non-equilibrium configuration
(i.e. we calculate $E=\sum_{\mathbf{q}}\epsilon_{\mathbf{q}}{}^{(f)}\, f_{F}(\epsilon_{\mathbf{q}}^{(i)},T_{i})$,
where $\epsilon_{\mathbf{q}}{}^{(i)}$ and $\epsilon_{\mathbf{q}}{}^{(f)}$
are the single particle energies for the initial and final lattice
depths respectively, and $f_{F}$ is the Fermi distribution function).
This procedure assumes that there has been no collisional redistribution
to allow the system to adjust to the lattice potential during the
period it is changed. To determine the thermodynamic state the final
distribution will relax to, we use the energy of the non-equilibrium
distribution as a constraint for finding the equilibrium values of
temperature and entropy. In general the final state properties will
depend on the initial temperature, filling factor, and final depth
of the lattice. To illustrate typical behavior we show a set of fast
loading process curves in Fig. \ref{FIG:nonadiab} for two different
values of the filling factor.

These curves show, as is expected from standard thermodynamic arguments,
that entropy increases for non-adiabatic processes, i.e. all loading
curves in Figs. \ref{FIG:nonadiab}(a) and (b) bend upwards with increasing
lattice depth. For the results with filling factor $n=0.25$ and for
initial temperatures deep in the cooling regime (i.e. initial states
far below the entropy plateau) a useful degree of temperature reduction
can be achieved with fast lattice loading up to certain maximum depth.
For example, the second lowest fast loading curve in Fig. \ref{FIG:nonadiab}(a)
cools with increasing lattice depth up to $V\approx15E_{R}$, and
then begins to heat for larger final lattice depths. Generally for
low filling factors ($n<1$) where the ground band plays the dominant
role in the system behavior at low temperatures, the entropy increase
is due to mainly to the reshaping of the single particle energy states
which occurs at low lattice depth %
\footnote{As the lattice is ramped up the free particle dispersion relation
$\epsilon(\mathbf{q})=\hbar^{2}\mathbf{q}^{2}/2m$ rapidly changes
to a tight-binding form $\epsilon_{{\textrm{TB}}}(\mathbf{q})$ (see
Eq. (\ref{eq:TB})). As the lattice depth increases further the ground
band energy states compress more (i.e. scale) but do not undergo further
reshaping.%
}. This effect can be reduced by taking as the initial condition for
the fast loading a system in equilibrium at a finite lattice depth
for which the dispersion relation is more tight-binding-like. This
situation was considered in Ref. \cite{Hofstetter2002} in application
to preparing a superfluid Fermi gas using in an optical lattice. Their
results, for the case $n=0.5$ and an initial lattice depth of $V\approx1E_{R}$,
predicted a useful degree of cooling.

As was demonstrated in Fig. \ref{FIG:ST1}(c), for filling factor
$n=1$ adiabatic lattice loading causes the atoms to heat. This effect
is exasperated by non-adiabatic loading, as shown in Fig. \ref{FIG:nonadiab}(b).
This case also benefits from beginning in a lattice of non-zero depth,
since at fixed temperature but increasing lattice depth (hence larger
$\epsilon_{{\textrm{BW}}}$), a smaller number of particles will be
found in the excited bands.

\section{Conclusion}

In this paper we have calculated the entropy-temperature curves for
fermions in a 3D optical lattice at various depths and filling factors.
We have identified general features of the thermodynamic properties
relevant to lattice loading, indicated regimes where adiabatically
changing the lattice depth will cause heating or cooling of the atomic
sample, and have provided limiting results for the behavior of the
entropy curves.The results presented in this work suggest optimal
regimes (filling factors and temperatures) which will facilitate the
suppression of thermal fluctuations in a fermionic gas by lattice
loading. These predictions should be easily verifiable with current
experiments. We have also shown that for a sample of fermions with
filling factor greater than one, the cooling regime is accompanied
by a significant reduction of the temperature compared to the Fermi
temperature. This regime would clearly be desirable for experiments
to investigate as an avenue for producing dilute Fermi gases with
$T/T_{F}\ll1$. We have shown that many of our predictions are robust
to non-adiabatic effects.

\section*{Acknowledgments}

PBB would like to thank C.W. Clark (NIST) for support during the initial
stages of this research. PBB would like to thank the referee for useful
suggestions related to the effects of lattice loading on $T/T_{F}$.

\bibliographystyle{/localscratch/bblakie/documents/PapersInProgress/projector/apsrev}
\bibliography{/localscratch/bblakie/documents/PapersInProgress/FermionicCooling/bandstructbib}

\begin{thebibliography}{25}
\expandafter\ifx\csname natexlab\endcsname\relax\def\natexlab#1{#1}\fi
\expandafter\ifx\csname bibnamefont\endcsname\relax
  \def\bibnamefont#1{#1}\fi
\expandafter\ifx\csname bibfnamefont\endcsname\relax
  \def\bibfnamefont#1{#1}\fi
\expandafter\ifx\csname url\endcsname\relax
  \def\url#1{\texttt{#1}}\fi
\expandafter\ifx\csname urlprefix\endcsname\relax\def\urlprefix{URL }\fi
\providecommand*{\bibinfo}[2]{#2}
\providecommand*{\eprint}[1]{#1}
\providecommand*{\url}[1]{#1}
\begingroup\makeatletter
 \@temptokena{%
  \expandafter\ifx\csname citenamefont\endcsname\relax
   \DeclareRobustCommand\citenamefont{\@firstofone}%
   \global\let\citenamefont\citenamefont
   \global\expandafter\let\csname citenamefont \expandafter\endcsname\csname
  citenamefont \endcsname
  \fi
 }\if@filesw\immediate\write\@auxout{\the\@temptokena}\fi
\expandafter\endgroup\the\@temptokena

\bibitem[{\citenamefont{DeMarco and Jin}(1999)}]{DeMarco1999a}
\bibinfo{author}{\bibfnamefont{B.}~\bibnamefont{DeMarco}} \bibnamefont{and}
  \bibinfo{author}{\bibfnamefont{D.~S.} \bibnamefont{Jin}},
  \bibinfo{journal}{Science}
  \textbf{\bibinfo{volume}{285}}(\bibinfo{number}{5434}), \bibinfo{pages}{1703}
  (\bibinfo{year}{1999}).

\bibitem[{\citenamefont{Schreck} \emph{et~al.}(2001)\citenamefont{Schreck,
  Khaykovich, Corwin, Ferrari, Bourdel, Cubizolles, and
  Salomon}}]{Schreck2001a}
\bibinfo{author}{\bibfnamefont{F.}~\bibnamefont{Schreck}},
  \bibinfo{author}{\bibfnamefont{L.}~\bibnamefont{Khaykovich}},
  \bibinfo{author}{\bibfnamefont{K.~L.} \bibnamefont{Corwin}},
  \bibinfo{author}{\bibfnamefont{G.}~\bibnamefont{Ferrari}},
  \bibinfo{author}{\bibfnamefont{T.}~\bibnamefont{Bourdel}},
  \bibinfo{author}{\bibfnamefont{J.}~\bibnamefont{Cubizolles}},
  \bibnamefont{and} \bibinfo{author}{\bibfnamefont{C.}~\bibnamefont{Salomon}},
  \bibinfo{journal}{Phys. Rev. Lett.}
  \textbf{\bibinfo{volume}{87}}(\bibinfo{number}{8}), \bibinfo{pages}{080403}
  (\bibinfo{year}{2001}).

\bibitem[{\citenamefont{O'Hara} \emph{et~al.}(2002)\citenamefont{O'Hara,
  Hemmer, Gehm, Granade, and Thomas}}]{O'Hara2002a}
\bibinfo{author}{\bibfnamefont{K.~M.} \bibnamefont{O'Hara}},
  \bibinfo{author}{\bibfnamefont{S.~L.} \bibnamefont{Hemmer}},
  \bibinfo{author}{\bibfnamefont{M.~E.} \bibnamefont{Gehm}},
  \bibinfo{author}{\bibfnamefont{S.~R.} \bibnamefont{Granade}},
  \bibnamefont{and} \bibinfo{author}{\bibfnamefont{J.~E.}
  \bibnamefont{Thomas}}, \bibinfo{journal}{Science}
  \textbf{\bibinfo{volume}{298}}, \bibinfo{pages}{2179} (\bibinfo{year}{2002}).

\bibitem[{\citenamefont{Modugno} \emph{et~al.}(2002)\citenamefont{Modugno,
  Roati, Riboli, Ferlaino, Brecha, and Inguscio}}]{Modugno2002a}
\bibinfo{author}{\bibfnamefont{G.}~\bibnamefont{Modugno}},
  \bibinfo{author}{\bibfnamefont{G.}~\bibnamefont{Roati}},
  \bibinfo{author}{\bibfnamefont{F.}~\bibnamefont{Riboli}},
  \bibinfo{author}{\bibfnamefont{F.}~\bibnamefont{Ferlaino}},
  \bibinfo{author}{\bibfnamefont{R.~J.} \bibnamefont{Brecha}},
  \bibnamefont{and} \bibinfo{author}{\bibfnamefont{M.}~\bibnamefont{Inguscio}},
  \bibinfo{journal}{Science}
  \textbf{\bibinfo{volume}{297}}(\bibinfo{number}{5590}), \bibinfo{pages}{2240}
  (\bibinfo{year}{2002}).

\bibitem[{\citenamefont{Gupta} \emph{et~al.}(2003)\citenamefont{Gupta,
  Hadzibabic, Zwierlein, Stan, Dieckmann, Schunck, van Kempen, Verhaar, and
  Ketterle}}]{Gupta2003a}
\bibinfo{author}{\bibfnamefont{S.}~\bibnamefont{Gupta}},
  \bibinfo{author}{\bibfnamefont{Z.}~\bibnamefont{Hadzibabic}},
  \bibinfo{author}{\bibfnamefont{M.~W.} \bibnamefont{Zwierlein}},
  \bibinfo{author}{\bibfnamefont{C.~A.} \bibnamefont{Stan}},
  \bibinfo{author}{\bibfnamefont{K.}~\bibnamefont{Dieckmann}},
  \bibinfo{author}{\bibfnamefont{C.~H.} \bibnamefont{Schunck}},
  \bibinfo{author}{\bibfnamefont{E.~G.~M.} \bibnamefont{van Kempen}},
  \bibinfo{author}{\bibfnamefont{B.~J.} \bibnamefont{Verhaar}},
  \bibnamefont{and} \bibinfo{author}{\bibfnamefont{W.}~\bibnamefont{Ketterle}},
  \bibinfo{journal}{Science} \textbf{\bibinfo{volume}{300}},
  \bibinfo{pages}{1723} (\bibinfo{year}{2003}).

\bibitem[{\citenamefont{Regal} \emph{et~al.}(2003)\citenamefont{Regal, Ticknor,
  Bohn, and Jin}}]{Regal2003a}
\bibinfo{author}{\bibfnamefont{C.~A.} \bibnamefont{Regal}},
  \bibinfo{author}{\bibfnamefont{C.}~\bibnamefont{Ticknor}},
  \bibinfo{author}{\bibfnamefont{J.~L.} \bibnamefont{Bohn}}, \bibnamefont{and}
  \bibinfo{author}{\bibfnamefont{D.~S.} \bibnamefont{Jin}},
  \bibinfo{journal}{Nature} \textbf{\bibinfo{volume}{424}}, \bibinfo{pages}{47}
  (\bibinfo{year}{2003}).

\bibitem[{\citenamefont{Cubizolles}
  \emph{et~al.}(2003)\citenamefont{Cubizolles, Bourdel, Kokkelmans,
  Shlyapnikov, and Salomon}}]{Cubizolle2003a}
\bibinfo{author}{\bibfnamefont{J.}~\bibnamefont{Cubizolles}},
  \bibinfo{author}{\bibfnamefont{T.}~\bibnamefont{Bourdel}},
  \bibinfo{author}{\bibfnamefont{S.~J. J. M.~F.} \bibnamefont{Kokkelmans}},
  \bibinfo{author}{\bibfnamefont{G.}~\bibnamefont{Shlyapnikov}},
  \bibnamefont{and} \bibinfo{author}{\bibfnamefont{C.}~\bibnamefont{Salomon}},
  \bibinfo{journal}{Phys. Rev. Lett.} \textbf{\bibinfo{volume}{91}},
  \bibinfo{pages}{240401} (\bibinfo{year}{2003}).

\bibitem[{\citenamefont{Hofstetter}
  \emph{et~al.}(2002)\citenamefont{Hofstetter, Cirac, Zoller, Demler, and
  Lukin}}]{Hofstetter2002}
\bibinfo{author}{\bibfnamefont{W.}~\bibnamefont{Hofstetter}},
  \bibinfo{author}{\bibfnamefont{J.~I.} \bibnamefont{Cirac}},
  \bibinfo{author}{\bibfnamefont{P.}~\bibnamefont{Zoller}},
  \bibinfo{author}{\bibfnamefont{E.}~\bibnamefont{Demler}}, \bibnamefont{and}
  \bibinfo{author}{\bibfnamefont{M.~D.} \bibnamefont{Lukin}},
  \bibinfo{journal}{Phys. Rev. Lett.} \textbf{\bibinfo{volume}{89}},
  \bibinfo{pages}{220407} (\bibinfo{year}{2002}).

\bibitem[{\citenamefont{Rabl} \emph{et~al.}(2003)\citenamefont{Rabl, Daley,
  Fedichev, Cirac, and Zoller}}]{Rabl2003}
\bibinfo{author}{\bibfnamefont{P.}~\bibnamefont{Rabl}},
  \bibinfo{author}{\bibfnamefont{A.~J.} \bibnamefont{Daley}},
  \bibinfo{author}{\bibfnamefont{P.~O.} \bibnamefont{Fedichev}},
  \bibinfo{author}{\bibfnamefont{J.~I.} \bibnamefont{Cirac}}, \bibnamefont{and}
  \bibinfo{author}{\bibfnamefont{P.}~\bibnamefont{Zoller}},
  \bibinfo{journal}{Phys. Rev. Lett.} \textbf{\bibinfo{volume}{91}},
  \bibinfo{pages}{110403} (\bibinfo{year}{2003}).

\bibitem[{\citenamefont{Viverit} \emph{et~al.}(2004)\citenamefont{Viverit,
  Menotti, Calarco, and Smerzi}}]{Viverit2004}
\bibinfo{author}{\bibfnamefont{L.}~\bibnamefont{Viverit}},
  \bibinfo{author}{\bibfnamefont{C.}~\bibnamefont{Menotti}},
  \bibinfo{author}{\bibfnamefont{T.}~\bibnamefont{Calarco}}, \bibnamefont{and}
  \bibinfo{author}{\bibfnamefont{A.}~\bibnamefont{Smerzi}},
  \bibinfo{journal}{Phys. Rev. Lett.} \textbf{\bibinfo{volume}{93}},
  \bibinfo{pages}{110401} (\bibinfo{year}{2004}).

\bibitem[{\citenamefont{Santos} \emph{et~al.}(2004)\citenamefont{Santos,
  Baranov, Cirac, Everts, Fehrmann, and Lewenstein}}]{Santos2004}
\bibinfo{author}{\bibfnamefont{L.}~\bibnamefont{Santos}},
  \bibinfo{author}{\bibfnamefont{M.~A.} \bibnamefont{Baranov}},
  \bibinfo{author}{\bibfnamefont{J.~I.} \bibnamefont{Cirac}},
  \bibinfo{author}{\bibfnamefont{H.-U.} \bibnamefont{Everts}},
  \bibinfo{author}{\bibfnamefont{H.}~\bibnamefont{Fehrmann}}, \bibnamefont{and}
  \bibinfo{author}{\bibfnamefont{M.}~\bibnamefont{Lewenstein}},
  \bibinfo{journal}{Phys. Rev. Lett.} \textbf{\bibinfo{volume}{93}},
  \bibinfo{pages}{030601} (\bibinfo{year}{2004}).

\bibitem[{\citenamefont{Modugno} \emph{et~al.}(2003)\citenamefont{Modugno,
  Ferlaino, Heidemann, Roati, and Inguscio}}]{Modugno2003a}
\bibinfo{author}{\bibfnamefont{G.}~\bibnamefont{Modugno}},
  \bibinfo{author}{\bibfnamefont{F.}~\bibnamefont{Ferlaino}},
  \bibinfo{author}{\bibfnamefont{R.}~\bibnamefont{Heidemann}},
  \bibinfo{author}{\bibfnamefont{G.}~\bibnamefont{Roati}}, \bibnamefont{and}
  \bibinfo{author}{\bibfnamefont{M.}~\bibnamefont{Inguscio}},
  \bibinfo{journal}{Phys. Rev. A} \textbf{\bibinfo{volume}{68}},
  \bibinfo{pages}{011601(R)} (\bibinfo{year}{2003}).

\bibitem[{\citenamefont{Ott} \emph{et~al.}(2004)\citenamefont{Ott, de~Mirandes,
  Ferlaino, Roati, Modugno, and Inguscio}}]{Ott2004a}
\bibinfo{author}{\bibfnamefont{H.}~\bibnamefont{Ott}},
  \bibinfo{author}{\bibfnamefont{E.}~\bibnamefont{de~Mirandes}},
  \bibinfo{author}{\bibfnamefont{F.}~\bibnamefont{Ferlaino}},
  \bibinfo{author}{\bibfnamefont{G.}~\bibnamefont{Roati}},
  \bibinfo{author}{\bibfnamefont{G.}~\bibnamefont{Modugno}}, \bibnamefont{and}
  \bibinfo{author}{\bibfnamefont{M.}~\bibnamefont{Inguscio}},
  \bibinfo{journal}{Phys. Rev. Lett.} \textbf{\bibinfo{volume}{92}},
  \bibinfo{pages}{160601} (\bibinfo{year}{2004}).

\bibitem[{\citenamefont{Orzel} \emph{et~al.}(2001)\citenamefont{Orzel, Tuchman,
  Fenselau, Yasuda, and Kasevich}}]{Orzel2001a}
\bibinfo{author}{\bibfnamefont{C.}~\bibnamefont{Orzel}},
  \bibinfo{author}{\bibfnamefont{A.~K.} \bibnamefont{Tuchman}},
  \bibinfo{author}{\bibfnamefont{M.~L.} \bibnamefont{Fenselau}},
  \bibinfo{author}{\bibfnamefont{M.}~\bibnamefont{Yasuda}}, \bibnamefont{and}
  \bibinfo{author}{\bibfnamefont{M.~A.} \bibnamefont{Kasevich}},
  \bibinfo{journal}{Science}
  \textbf{\bibinfo{volume}{23}}(\bibinfo{number}{291}), \bibinfo{pages}{2386}
  (\bibinfo{year}{2001}).

\bibitem[{\citenamefont{Greiner}
  \emph{et~al.}(2002{\natexlab{a}})\citenamefont{Greiner, Mandel, H{\"a}nsch,
  and Bloch}}]{Greiner2002b}
\bibinfo{author}{\bibfnamefont{M.}~\bibnamefont{Greiner}},
  \bibinfo{author}{\bibfnamefont{O.}~\bibnamefont{Mandel}},
  \bibinfo{author}{\bibfnamefont{T.~W.} \bibnamefont{H{\"a}nsch}},
  \bibnamefont{and} \bibinfo{author}{\bibfnamefont{I.}~\bibnamefont{Bloch}},
  \bibinfo{journal}{Nature} \textbf{\bibinfo{volume}{419}}, \bibinfo{pages}{51}
  (\bibinfo{year}{2002}{\natexlab{a}}).

\bibitem[{\citenamefont{Greiner}
  \emph{et~al.}(2002{\natexlab{b}})\citenamefont{Greiner, Mandel, Esslinger,
  H{\"a}nsch, and Bloch}}]{Greiner2002a}
\bibinfo{author}{\bibfnamefont{M.}~\bibnamefont{Greiner}},
  \bibinfo{author}{\bibfnamefont{O.}~\bibnamefont{Mandel}},
  \bibinfo{author}{\bibfnamefont{T.}~\bibnamefont{Esslinger}},
  \bibinfo{author}{\bibfnamefont{T.~W.} \bibnamefont{H{\"a}nsch}},
  \bibnamefont{and} \bibinfo{author}{\bibfnamefont{I.}~\bibnamefont{Bloch}},
  \bibinfo{journal}{Nature} \textbf{\bibinfo{volume}{415}}, \bibinfo{pages}{39}
  (\bibinfo{year}{2002}{\natexlab{b}}).

\bibitem[{\citenamefont{Mandel} \emph{et~al.}(2003)\citenamefont{Mandel,
  Greiner, Widera, Rom, H{\"a}nsch, and Bloch}}]{Mandel2003a}
\bibinfo{author}{\bibfnamefont{O.}~\bibnamefont{Mandel}},
  \bibinfo{author}{\bibfnamefont{M.}~\bibnamefont{Greiner}},
  \bibinfo{author}{\bibfnamefont{A.}~\bibnamefont{Widera}},
  \bibinfo{author}{\bibfnamefont{T.}~\bibnamefont{Rom}},
  \bibinfo{author}{\bibfnamefont{T.}~\bibnamefont{H{\"a}nsch}},
  \bibnamefont{and} \bibinfo{author}{\bibfnamefont{I.}~\bibnamefont{Bloch}},
  \bibinfo{journal}{Phys. Rev. Lett.} \textbf{\bibinfo{volume}{91}},
  \bibinfo{pages}{010407} (\bibinfo{year}{2003}).

\bibitem[{\citenamefont{Rom} \emph{et~al.}(2004)\citenamefont{Rom, Best,
  Mandel, Widera, Greiner, H{\"a}nsch, and Bloch}}]{Rom2004a}
\bibinfo{author}{\bibfnamefont{T.}~\bibnamefont{Rom}},
  \bibinfo{author}{\bibfnamefont{T.}~\bibnamefont{Best}},
  \bibinfo{author}{\bibfnamefont{O.}~\bibnamefont{Mandel}},
  \bibinfo{author}{\bibfnamefont{A.}~\bibnamefont{Widera}},
  \bibinfo{author}{\bibfnamefont{M.}~\bibnamefont{Greiner}},
  \bibinfo{author}{\bibfnamefont{T.~W.} \bibnamefont{H{\"a}nsch}},
  \bibnamefont{and} \bibinfo{author}{\bibfnamefont{I.}~\bibnamefont{Bloch}},
  \bibinfo{journal}{Phys. Rev. Lett.} \textbf{\bibinfo{volume}{93}},
  \bibinfo{pages}{073002} (\bibinfo{year}{2004}).

\bibitem[{\citenamefont{Kastberg} \emph{et~al.}(1995)\citenamefont{Kastberg,
  Phillips, Rolston, Spreeuw, and Jessen}}]{Kastberg1995a}
\bibinfo{author}{\bibfnamefont{A.}~\bibnamefont{Kastberg}},
  \bibinfo{author}{\bibfnamefont{W.~D.} \bibnamefont{Phillips}},
  \bibinfo{author}{\bibfnamefont{S.~L.} \bibnamefont{Rolston}},
  \bibinfo{author}{\bibfnamefont{R.~J.~C.} \bibnamefont{Spreeuw}},
  \bibnamefont{and} \bibinfo{author}{\bibfnamefont{P.~S.}
  \bibnamefont{Jessen}}, \bibinfo{journal}{Phys. Rev. Lett.}
  \textbf{\bibinfo{volume}{74}}, \bibinfo{pages}{1542} (\bibinfo{year}{1995}).

\bibitem[{\citenamefont{Blakie and Porto}(2004)}]{Blakie2004a}
\bibinfo{author}{\bibfnamefont{P.~B.} \bibnamefont{Blakie}} \bibnamefont{and}
  \bibinfo{author}{\bibfnamefont{J.~V.} \bibnamefont{Porto}},
  \bibinfo{journal}{Phys. Rev. A} \textbf{\bibinfo{volume}{69}},
  \bibinfo{pages}{013603} (\bibinfo{year}{2004}).

\bibitem[{\citenamefont{Jaksch} \emph{et~al.}(1998)\citenamefont{Jaksch,
  Bruder, Cirac, Gardiner, and Zoller}}]{Jaksch1998a}
\bibinfo{author}{\bibfnamefont{D.}~\bibnamefont{Jaksch}},
  \bibinfo{author}{\bibfnamefont{C.}~\bibnamefont{Bruder}},
  \bibinfo{author}{\bibfnamefont{J.~I.} \bibnamefont{Cirac}},
  \bibinfo{author}{\bibfnamefont{C.}~\bibnamefont{Gardiner}}, \bibnamefont{and}
  \bibinfo{author}{\bibfnamefont{P.}~\bibnamefont{Zoller}},
  \bibinfo{journal}{Phys. Rev. Lett.} \textbf{\bibinfo{volume}{81}},
  \bibinfo{pages}{3108} (\bibinfo{year}{1998}).

\bibitem[{\citenamefont{van Oosten} \emph{et~al.}(2001)\citenamefont{van
  Oosten, van~der Straten, and Stoof}}]{Oosten2001a}
\bibinfo{author}{\bibfnamefont{D.}~\bibnamefont{van Oosten}},
  \bibinfo{author}{\bibfnamefont{P.}~\bibnamefont{van~der Straten}},
  \bibnamefont{and} \bibinfo{author}{\bibfnamefont{H.~T.~C.}
  \bibnamefont{Stoof}}, \bibinfo{journal}{Phys. Rev. A}
  \textbf{\bibinfo{volume}{63}}, \bibinfo{pages}{053601}
  (\bibinfo{year}{2001}).

\bibitem[{\citenamefont{Sklarz} \emph{et~al.}(2002)\citenamefont{Sklarz,
  Friedler, Tannor, Band, and Williams}}]{Sklarzr2002a}
\bibinfo{author}{\bibfnamefont{S.~E.} \bibnamefont{Sklarz}},
  \bibinfo{author}{\bibfnamefont{I.}~\bibnamefont{Friedler}},
  \bibinfo{author}{\bibfnamefont{D.~J.} \bibnamefont{Tannor}},
  \bibinfo{author}{\bibfnamefont{Y.~B.} \bibnamefont{Band}}, \bibnamefont{and}
  \bibinfo{author}{\bibfnamefont{C.~J.} \bibnamefont{Williams}},
  \bibinfo{journal}{Phys. Rev. A} \textbf{\bibinfo{volume}{66}},
  \bibinfo{pages}{053620} (\bibinfo{year}{2002}).

\bibitem[{\citenamefont{Denschlag} \emph{et~al.}(2002)\citenamefont{Denschlag,
  Simsarian, H\"{a}ffner, McKenzie, Browaeys, Cho, Helmerson, Rolston, and
  Phillips}}]{Denschlag2002a}
\bibinfo{author}{\bibfnamefont{J.~H.} \bibnamefont{Denschlag}},
  \bibinfo{author}{\bibfnamefont{J.~E.} \bibnamefont{Simsarian}},
  \bibinfo{author}{\bibfnamefont{H.}~\bibnamefont{H\"{a}ffner}},
  \bibinfo{author}{\bibfnamefont{C.}~\bibnamefont{McKenzie}},
  \bibinfo{author}{\bibfnamefont{A.}~\bibnamefont{Browaeys}},
  \bibinfo{author}{\bibfnamefont{D.}~\bibnamefont{Cho}},
  \bibinfo{author}{\bibfnamefont{K.}~\bibnamefont{Helmerson}},
  \bibinfo{author}{\bibfnamefont{S.~L.} \bibnamefont{Rolston}},
  \bibnamefont{and} \bibinfo{author}{\bibfnamefont{W.~D.}
  \bibnamefont{Phillips}}, \bibinfo{journal}{J. Phys. B}
  \textbf{\bibinfo{volume}{35}}(\bibinfo{number}{14}), \bibinfo{pages}{3095}
  (\bibinfo{year}{2002}).

\bibitem[{\citenamefont{Ashcroft and Mermin}(1976)}]{Mermin1976}
\bibinfo{author}{\bibfnamefont{N.~W.} \bibnamefont{Ashcroft}} \bibnamefont{and}
  \bibinfo{author}{\bibfnamefont{N.~D.} \bibnamefont{Mermin}},
  \emph{\bibinfo{title}{Solid State Physics}} (\bibinfo{publisher}{W.B.
  Saunders Company}, \bibinfo{year}{1976}).

\end{thebibliography}

\end{document}